\documentclass{JHEP3}

\usepackage{graphicx}

\title {Counting Supertubes}

\author{Belkis Cabrera Palmer\\
Physics Department, Syracuse University, Syracuse, NY 13244,\\
Physics Department, UCSB, Santa Barbara, CA 93106.
\texttt{bcabrera@physics.ucsb.edu}}

\author{Donald Marolf\\
Physics Department, UCSB, Santa Barbara, CA 93106.
\texttt{marolf@physics.ucsb.edu}}

\abstract{The quantum states of the supertube are counted by
directly quantizing the linearized Born-Infeld action near the
round tube. The result is an entropy $S = 2\pi \sqrt{2
(Q_{D0}Q_{F1}-J)}$, in accord with conjectures in the literature.
As a result, supertubes may be the generic D0-F1 bound state. Our
approach also shows directly that supertubes are marginal bound
states with a discrete spectrum. We also discuss the relation to
recent suggestions of Mathur et al involving  three-charge black
holes.}

\date{February, 2004}

\preprint{hep-th/0403025} 

\begin{document}

\section{Introduction}

Two hallmarks of modern stringy physics are the non-abelian
interactions associated with multiple branes, and the idea that
such non-abelian behavior can simplify and dramatically change
form in the large $N$ limit.  Gauge/gravity dualities
\cite{BFSS,Juan,ISMY,MAGOO} and the related stringy counting of
black hole entropy \cite{SV} are examples where there seems to be
a strict duality at large $N$. However, many other interesting
examples arise in brane polarization effects (see e.g.
\cite{emparan,myers}) where a bound state of low-dimension branes
may, when polarized, be effectively described as a single brane of
higher dimension.

We will be concerned here with the D0-F1 supertube of
\cite{MateosTownsend,EmparanMateosTownsend,Ohta}, which is an example
of such brane polarization.  Supertubes have the special
distinction that the polarized states are BPS and arise without
the application of an external field.  Mathur et al
\cite{LuninMathur1,LuninMathur2,MathurEssay,Mathur3Q} have
suggested that supertubes are also connected to black hole entropy
(and thus to $AdS_3/CFT_2$).

The supertubes of interest here carry D0 and F1 charge and have
the supersymmetries expected of such configurations.  These are
the original supertubes of \cite{MateosTownsend}, though many
related configurations can be obtained through duality
transformations. The charges are arrayed around a tube of topology
$S^1 \times \mathbb{R}$ in space, where the $ \mathbb{R}$
represents a translation symmetry of the system and the direction
along which the fundamental strings are aligned. Interestingly,
the $S^1$ can be an arbitrary curve \cite{MateosNgTownsend} (see also \cite{BL,BaKa,BO,BOetal} for earlier and related results) in the
space of symmetry orbits; all such configurations are static.  We
assume the curve is closed and also compactify the $\mathbb{R}$
direction as we are interested in cases with finite charge.

Mateos and Townsend \cite{MateosTownsend} showed that the
supertube can be described using the Dirac-Born-Infeld effective
action of a D2-brane. The D2-worldvolume is then the
above-mentioned $S^1 \times \mathbb{R}$ tube ($\times$ time).
Because the $S^1$ is a closed curve, the configuration has no net
D2 charge. However, if the U(1) electric and magnetic fields ($E$
and $B$) are switched on, the configuration gains both a net D0
and a net F1 charge. Supertubes arise when the electric field
reaches $E=1$ in string units (with $2\pi \alpha' =1$) and when
$B$ is nowhere vanishing. The static nature of the supertube can
be understood as a balance between the D2-brane tension and the
Poynting angular momentum from the simultaneous presence of both
electric and magnetic fields
\cite{MateosTownsend,MateosNgTownsend}.

It is natural to conjecture \cite{MateosTownsend} that supertubes
describe D0-F1 bound states.  Because they would be marginal such
states, it is nontrivial to verify that they are in fact bound.
However, we will demonstrate this in section \ref{ops} through an
explicit quantization of the system in which the spectrum of BPS
states is shown to be discrete\footnote{We will use a linearized
description in which the DeWit-Hoppe-Nicolai continuum of membrane
states \cite{DeWitHoppeNicolai} does not arise.  This is
consistent \cite{BFSS} with our intent to study a single bound
state, and not the second-quantized theory of supertubes.}.

A much stronger conjecture is that {\it almost all} D0-F1 bound
states are supertubes for large $Q_{D0},Q_{F1},J$.  This would be
of great interest, as supertubes would then provide a {\it
geometric} description of these bound states; states of the
supertube are directly labelled by the shape of the $S^1$
cross-section and by the magnetic field as a function of location
on this $S^1$.  In contrast, previously known descriptions
are highly non-geometric; the states are only indirectly understood in terms of
the non-abelian D0-F1 gauge theory.  Mathur et al have provided evidence that this conjecture is correct, but an explicit quantization and counting has remained lacking.  Mathur et al have also made
interesting further conjectures extrapolating to the
three-charge systems associated with black holes, but we will save discussion of such
conjectures for section \ref{disc}.

An intermediate conjecture was made in \cite{LuninMaldacenaMaoz} to
the effect that supertubes are in fact a significant fraction of
such BPS D0-F1 bound states. In particular, Lunin, Maldacena, and Maoz
conjectured that the entropy of a supertube configuration
with $Q_{D0}$ units of D0 charge and $Q_{F1}$ units of F1 charge
 is $S = 2\pi \sqrt{2 Q_{D0}Q_{F1}}$ to leading order in large charges.  Note that this is just the leading
order expression for the entropy of all BPS D0-F1 bound states,
which can be computed from the fact that the system is dual to the
fundamental string with right-moving momentum, whose entropy is in
turn given by the Cardy formula \cite{Cardy}.  It is also of
interest to count supertubes with fixed angular momentum $J$, in
which case the corresponding conjecture would be $S = 2\pi \sqrt{2
(Q_{D0}Q_{F1}-J)}$ \cite{LuninMathur2} . The main point of our
work below is to verify this conjecture by using a linearization
of the D2-brane effective action to directly count quantum
supertube states.

The structure of this paper is as follows.  We begin with some
preliminaries in section \ref{prelim}, gauge fixing the D2
effective action and linearizing about the round supertube
configuration (in which the $S^1$ is a circle).  However, we save
the general justification of certain interesting relations for
Appendix A and the detailed form of certain expressions for
Appendix B.  The spectrum of states is then computed in section
\ref{spectrum}, whence it is straightforward to count the states
and to establish that our results are valid when $Q_{D0}Q_{F1} -J
\ll Q_{D0}Q_{F1}$ and $Q_{D0}Q_{F1} -J \gg 1$. As stated above,
our counting verifies that supertubes are marginal bound states
with an entropy given by $S = 2\pi \sqrt{2 (Q_{D0}Q_{F1}-J)}$.

Finally, we close in section \ref{disc} with a summary and a
discussion of the implications for the further conjectures of
Mathur et al \cite{LuninMathur1,LuninMathur2,MathurEssay,Mathur3Q}
relating to three-charge black holes.  While the D2 action used here
is not sufficient to describe the relevant three-charge
systems\footnote{See \cite{BK} for a Born-Infeld discussion of
three-charge supertubes.}, one expects the three-charge
calculations to proceed along similar lines.

\section{Preliminaries: Gauge Fixing and Linearization}
\label{prelim}

As indicated in the introduction, our starting point will be the
D2 Born-Infeld effective action:

\begin{equation}
\label{D2} S_{D2} =- T_{D2}  \int d^3 \xi \sqrt{-\det(g+{\cal F}))}
- T_{D2} \int C_{[1]} \wedge {\cal F},
\end{equation}
where ${\cal F_{\mu\nu}} = F_{\mu\nu} + B_{\mu\nu}$ and we have
included the Chern-Simons term representing the coupling to a
background Ramond-Ramond vector potential $C_{[1]}$ and a
Neveu-Schwarz two-form potential $B_{\mu\nu}$,  though we will
soon set all background fields to zero. $T_{D2}$ is the D2-brane
tension.

Our tasks are to isolate the physical degrees of freedom and to
find a description in which the states can be counted.  To this
end we impose a version of static gauge and then expand the action
and all relevant quantities to quadratic order in fields, taking
the round supertube (for which the $S^1$ is an isometry direction)
as the base point of the expansion.  Note
\cite{EmparanMateosTownsend,MateosNgTownsend} that this is the
unique configuration saturating the angular momentum bound on
supertubes.  As a result, it will have certain nice properties
reminiscent of vacuum states.

We now set all closed string fields to zero, e.g., we take the
supertube to be embedded in Minkowski space, and we set
$B_{\mu\nu} = 0$ and $C_{[1]} = 0$.  It will be convenient to
rename the worldvolume coordinates $\xi^0=t$, $\xi^1 = \sigma$,
$\xi^2 =z$. Taking $X^\mu$ to be Cartesian coordinates on
Minkowski space, our static gauge is then defined by the
conditions $t=X^0$, $z = X^3$, and $\tan \sigma = X^1/X^2$.  Thus,
$z$ represents the coordinate along the length of the tube, while
$\sigma$ is an angular coordinate around each $S^1$ cross-section
of the tube.  It is also convenient to introduce the radius
$R(t,z,\sigma)$ in the $X^1X^2$ plane defined by $R^2 = (X^1)^2 +
(X^2)^2$ and the notation $E = F_{tz}$, $B = F_{z \sigma}$.
Finally, we choose an electromagnetic gauge in which the
world-volume vector potential $A$ satisfies $A_t=0$.  In
particular, we take $A = Et dz + Bz d \sigma$ for the background
supertube.

In the usual way, the action (\ref{D2}) shows that the D2 brane
carries both D0 and F1 charge (in the $z$ direction). We are most interested in
static D2 branes invariant under $z$-translations, as adding time dependence or $z$-momentum breaks all supersymmetries.
The charges for this special case are given  (\cite{MateosNgTownsend}) by

\begin{eqnarray}
\label{charges0}
Q_{D0} &=& \frac{T_{D2}}{T_{D0}}\int dz d\sigma B  ,\\
\label{charges1} Q_{F1} &=& \frac{1}{T_{F1}} \int d\sigma \Pi_{z}
= \frac{T_{D2}}{T_{F1}} \int d\sigma \frac{E |\partial_\sigma
X|^2}{\sqrt{(1-E^2)|\partial_\sigma X|^2 + B^2}},
\end{eqnarray}
where $T_{D0},T_{F1}$ represent the tensions of the appropriate branes and we have normalized $Q_{D0}, Q_{F1}$ so that they take integer values.
The angular momentum $J$ (in the $X^1X^2$ plane) and energy
$P^0$ take the form
\begin{eqnarray}
J &=& T_{D2} \int d\sigma \frac{EB(X_1\partial_\sigma X_2-X_2\partial_\sigma X_1)}{\sqrt{(1-E^2)|\partial_\sigma X|^2 + B^2}} ,\\
P^0 &=& \int dz d \sigma {\cal T}^{00}(X(\xi))= T_{D2} \int dz d\sigma \frac{B^2+
|\partial_\sigma X|^2}{\sqrt{(1-E^2)|\partial_\sigma X|^2 + B^2}},
\end{eqnarray}
where ${\cal T}^{00} = \frac{2}{\sqrt{- G}} \frac{\partial L}{\partial G_{00}}$
is the stress-energy tensor on the D2-brane. The energy and charges satisfy a BPS bound \cite{MateosNgTownsend} of the form

\begin{equation}
\label{BPSbound} P^0 \ge T_{D0}| Q_{D0}| + T_{F1}|Q_{F1}| .
\end{equation}
We emphasize that the above charges are defined by the coupling of the D2 action (\ref{D2}) to
the metric and external gauge fields.

The round supertube is then given \cite{MateosTownsend} by
\begin{eqnarray}
 && R_{round}(t,z,\sigma) = R \ \ , \\
&& X^i_{round}(t,z,\sigma) = 0 \ \ {\rm for} \ \ i=4,5,6,7,8,9 \ \ , \\
&& (F_{tz})_{round} = \pm 1, \ \ (F_{z \sigma})_{round} = B \ \ ,
\end{eqnarray}
where, from here on, $R$ and $B$ are constants that determine the
charges and angular momentum of the round supertube about which we
expand below.

In order to make these charges finite, let us periodically
identify the system under $z \rightarrow z + L_z$, so that the
supertube at any time is an $S^1 \times S^1$ embedded in $S^1
\times \mathbb{R}^9$.  The total charges and angular momentum of
the round tube are then
\begin{eqnarray}
Q_{D0}^{round} &=& \frac{2\pi L_z T_{D2}}{T_{D0}} B , \\  Q_{F1}^{round} &=&{\rm sgn}(EB) \frac{2 \pi T_{D2}} {T_{F1}} \frac{R^2}{B} , \\   J^{round} &=& {\rm sgn}(EB)2 \pi L_z T_{D2} R^2 \ \ .
\end{eqnarray}
Note in particular that since $T_{D0} T_{F1} = 2 \pi T_{D2}$ (see, e.g. \cite{Joe}), we have
$J^{round} = Q_{D0}^{round} Q_{F1}^{round}$.

Our final task is to expand the action (\ref{D2}) to quadratic order
about the round supertube solution.  Let us denote the deviations
from the round solution by
\begin{eqnarray}
R &=& R_{round} + r, \ X^i = X^i_{round} + \eta^i, \\
A &=& A_{round} + a, \ F_{tz} = E_{round} + e_z, \ F_{z \sigma} = B_{round} +b \  \ {\rm and} \ \  F_{t \sigma} = e_\sigma.
\end{eqnarray}
It is then straightforward but tedious to expand the quantities of
interest to quadratic order in $\eta,a$.  Note that we also wish to compute the Hamiltonian $H = p\dot{q}- L $ associated with the resulting quadratic action (\ref{action}).    A general argument presented in Appendix A
shows that our gauge choice and properties of the Dirac-Born-Infeld action act together to guarantee that $H$ takes the form
\begin{equation}
\label{HBPS} H = P^0 - |Q_{D0}|T_{D0} - |Q_{F1}|T_{F1} L_z.
\end{equation}
In particular, (\ref{HBPS}) is not the energy of the system.  Instead, our Hamiltonian measures the extent to which a state
is excited above the BPS bound (\ref{BPSbound}).

The detailed results of the expansions in powers of fields are
useful for the next section, but are not particularly enlightening
in themselves. We will not burden the reader with such formulae here,
reserving them instead for Appendix B.  We will, however, mention
that in  computing the quadratic action (\ref{action}) from
(\ref{D2}), we perform an integration by parts so that the
canonical momenta take a more transparent form.  We warn the
reader in advance that this integration by parts performs a
canonical transformation that causes the canonical momentum
$\pi_z$ (conjugate to $a_z$) below to differ by linear terms from
the $\Pi_z$ (\ref{charges1}) defined from the original action
(\ref{D2}).  Thus, while the electric charge $Q_{F1}$ remains the
integral of $\Pi_z$, it is not the integral of the $\pi_z$ used
below.

\section{The spectrum of states}
\label{spectrum}
\label{ops}

We now use the results of section \ref{prelim} to find the
spectrum of states for our linearized system.  In fact, we can simplify the treatment
somewhat by realizing the momentum in the $z$ direction breaks supersymmetry. Since we
are interested in BPS states, we may thus restrict attention to modes independent of $z$.  The action for such modes is given in (\ref{action}), but the resulting equations of motion are:

\begin{eqnarray}
&& \frac{R^2+B^2}{B} \partial_t^2 r  +  {\rm sgn}(E)2(\partial_t \partial_{\sigma} r - \frac{R}{B}\partial_t a_z)=0 \\
&& \frac{R^2(R^2+B^2)}{B^3} \partial_t^2 a_z +{\rm sgn}(E)(\frac{2R^2}{B^2}\partial_t \partial_{\sigma} a_z + \frac{2R}{B} \partial_t r)=0, \\
&&\frac{R^2+B^2}{B} \partial_t^2 \eta^i  +{\rm sgn}(E)2\partial_t \partial_{\sigma} \eta^i =0, \ \ \  {\rm and} \\
&& \frac{1}{B}\partial_t^2 a_{\sigma}=0.
\end{eqnarray}
Note in particular that these equations are identically satisfied
when all time derivatives vanish, so that all static
configurations are allowed. We must also consider the Gauss Law
constraint which due to gauge fixing no longer follows from our
action.  However, for $z$-independent modes in our gauge this is
just $\partial_\sigma a_\sigma=0$ at this order.

Without loss of generality, we choose $ {\rm sgn}(E)={\rm sgn}(B)=1$.  We first compute the relevant mode expansions in section (\ref{mode}) and then count the relevant states in section (\ref{entropy}).

\subsection{Mode Expansions}
\label{mode}

Each transverse degree of freedom $\eta^i$ (for $i \in \{4,..9\})$ decouples from all other fields and has a solution of the form
\begin{eqnarray}
\label{etasols}
\eta^i &=&\frac{1}{ \sqrt{4 \pi L_z T_{D2}}} \sum_{k_\sigma \neq 0}\frac{ a^i_{k_\sigma}}{\sqrt{|k_\sigma|}} e^{i\omega_at+ik_\sigma \sigma}+\frac{b^i_{k_\sigma}}{\sqrt{|k_\sigma|}} e^{ik_\sigma \sigma},
\end{eqnarray}
where the normalizations have been chosen with foresight to
simplify expressions to come.  The relevant
frequencies are
\begin{equation}
\label{eadisperse}
\omega_a(k_\sigma) = -\frac{2Bk_\sigma}{R^2+B^2},\ \  \ \ {\rm and}\ \ \ \ \omega_b(k_\sigma) = 0.
\end{equation}

On the other hand, the radial and Maxwell degrees of freedom are coupled.  Their solutions take
the slightly more complicated form
\begin{eqnarray}
\label{sols}
r &=& \frac{1}{2 \sqrt{2 \pi L_z T_{D2}}}\sum_{k_{\sigma}\neq \pm 1} \frac{a^{\pm}_{k_{\sigma}}}{\sqrt{|-k_{\sigma}\pm 1|}} e^{i\omega_a^{\pm}t+ik_{\sigma} \sigma}+\frac{b^{\pm}_{k_{\sigma}}}{\sqrt{|-k_{\sigma} \pm 1|}} e^{ik_{\sigma} \sigma} ,\\
 a_z&=& \pm i\frac{B}{2R \sqrt{ 2\pi L_z T_{D2}}}\sum_{k_{\sigma}\neq \pm 1}\frac{a^{\pm}_{k_{\sigma}}}{\sqrt{|-k_{\sigma}\pm 1|}} e^{i\omega_a^{\pm}t+ik_{\sigma} \sigma}+\frac{b^{\pm}_{k_{\sigma}}}{\sqrt{|-k_{\sigma}  \pm 1|}} e^{ik_{\sigma} \sigma },\\
a_{\sigma}&=&   (const_1) t  +  const_2,
\end{eqnarray}
 with the similar but slightly more complicated frequencies
\begin{eqnarray}
\omega^\pm_a(k_\sigma) &=& \frac{2B}{R^2+B^2}(-k_{\sigma}\pm 1), \ \ {\rm and}\\
\omega^\pm_b(k_\sigma) &=& 0.
\end{eqnarray}
The $a_\sigma$ degree of freedom will not be of further interest below.

Note in particular that $\omega^\pm_a(k_\sigma)$ vanishes when $k
= \pm 1$.  These zero modes represent the translation symmetries
in the $X^1$ and $X^2$ direction. After quantization, such modes
become analogues of the free non-relativistic particle.  The same
is true of the $\eta^i$ modes with $k_\sigma=0$, associated with
translations in $X^i$ for $i\in \{4,...9\}$.  A careful treatment
shows that their velocities appear in the Hamiltonian $H$, so that
these modes are not annihilated by $H$ even though they have zero
frequency.  In particular, these modes are not BPS.  We will not
concern ourselves with the detailed treatment of these zero modes
here -- the expressions below should be understood as correct only
up to terms involving such modes.

In addition, we have $\omega_b^\pm (k_\sigma)= \omega_b(k_\sigma)
=0$ for all $k_\sigma$. This is just the linearized description of
the known result \cite{MateosNgTownsend} that the supertube allows
arbitrary static deformations of its cross-section and magnetic
field, so long as translation invariance in the $z$-direction is
preserved.   Although they have zero frequency, we will see below
that such modes are {\it not} described by free particle degrees
of freedom.  Instead, the coefficients
$a^\pm_{k_\sigma}$,$a^i_{k_\sigma}$ and
$b^\pm_{k_\sigma}$,$b^i_{k_\sigma}$ are standard creation and
annihilation operators which create or annihilate excitations of
the round supertube.  As a result, their vanishing frequency means
that these modes {\it are} annihilated by the linearized
Hamiltonain $H$. Since $H$ encodes the BPS condition, it is clear
that any $k_z=0$ excitation of the $b$-modes preserves the
BPS-bound.

>From the action (\ref{D2}) and the solutions (\ref{sols}), the
canonical momenta $\pi_z$ (conjugate to $a_z$) and ${\cal P}_r$,
${\cal P}_i$ take the form
\begin{eqnarray}
\label{momenta}
{\cal P}_r &=& -\frac{i}{2}\sqrt{\frac{ L_zT_{D2}}{2 \pi}} \sum_{k_\sigma \neq \pm 1} \frac{-k_\sigma\pm 1}{\sqrt{|-k_\sigma \pm 1|}}(a^\pm e^{i\omega_a^\pm t+ik_\sigma \sigma}-b^\pm e^{ik_\sigma \sigma}) ,\\
\pi_z&=& \pm \frac{R}{2B} \sqrt{\frac{L_zT_{D2}}{2 \pi}} \sum_{k_\sigma\neq \pm 1}\frac{-k_\sigma\pm 1}{\sqrt{|-k_\sigma \pm 1|}}(a^\pm e^{i\omega_a^\pm t +ik_\sigma \sigma}- b^\pm e^{ik_\sigma \sigma }), \\
{\cal P}_i &=& \sqrt{\frac{L_zT_{D2}}{4 \pi}} \sum_{k_\sigma \neq 0} \frac{-k_\sigma}{\sqrt{|k_\sigma |}}(a^i e^{i\omega_a t+ik_\sigma \sigma}-b^i e^{ik_\sigma \sigma}).
\end{eqnarray}

A straightforward but lengthy calculation from the canonical
commutation relations then shows that the $a$'s and
$b$'s  satisfy
\begin{eqnarray}
 && [a^+_{k_\sigma},a^-_{k'_\sigma}] =- \delta_{k_\sigma +k'_\sigma} {\rm sgn} (k_\sigma-1) ,\\
&& [b^+_{k_\sigma},b^-_{k'_\sigma}] = \delta_{k_\sigma
+k'_\sigma} {\rm sgn} (k_\sigma-1) ,\\
&& [a^i_{k_\sigma},a^i_{k'_\sigma}] = -\delta_{k_\sigma +k'_\sigma} {\rm sgn} (k_\sigma) ,\\
&& [b^i_{k_\sigma},b^i_{k'_\sigma}] = \delta_{k_\sigma
+k'_\sigma} {\rm sgn} (k_\sigma),
\end{eqnarray}
while the remaining commutators vanish. In addition, the reality conditions require
\begin{eqnarray}
&& (a^+_{k_\sigma})^\dagger = a^-_{-k_\sigma} ,\ \ \ (b^+_{k_\sigma})^\dagger = b^-_{-k_\sigma} ,\\
&& (a^i_{k_\sigma})^\dagger = a^i_{-k_\sigma} ,\ \ \ (b^i_{k_\sigma})^\dagger = b^i_{-k_\sigma}.
\end{eqnarray}

Thus we may identify  ($a^+_{k_\sigma}$,$b^-_{-k_\sigma}$)  for
$k_\sigma
> 1$  and  ($a^-_{-k_\sigma}$,$b^+_{k_\sigma}$) for $k_\sigma < 1$ as
creation operators and their adjoints as annihilation operators. Similarly, ($a^i_{k_\sigma}$,$b^i_{-k_\sigma}$) for $k_\sigma > 0$ are the creation operators for the $\eta$-modes.
In particular, for $k_z=0$ the BPS ($b$) modes carry negative
angular momentum around the cylinder while the non-BPS ($a$) modes
carry positive angular momentum.  This is in accord with the
result of \cite{MateosNgTownsend} that the round supertube is the
unique BPS state of maximal angular momentum.  As a result, the
round state acts like a vacuum state relative to the set of BPS
excitations\footnote{With the understanding that ``excitations"
{\it lower} the angular momentum instead of raising it.}.

Finally, we wish to express the charges in terms of the creation
and annihilation operators $a^\pm_{k_\sigma}, a^i_{k_\sigma}$ and $b^\pm_{k_\sigma}, b^i_{k_\sigma}$.
Once again, the procedure is straightforward but lengthy.  The
resulting expressions are:
\begin{eqnarray}
\label{opcharges}
 H &=&    \sum_{k_\sigma > 1} \omega^-_a(-k_\sigma) a^+_{k_\sigma} a^-_{-k_\sigma}+\sum_{k_\sigma < 1} \omega^+_a(k_\sigma) a^-_{-k_\sigma} a^+_{k_\sigma} + \sum_{k_\sigma > 0} \omega_a(k_\sigma) a^i_{k_\sigma} a^i_{-k_\sigma}  ,\\
%Q_{F1} &=& Q_{F1}^{round} +   \frac{2\pi T_{D2}}{T_{F1}}  \frac{R}{B} (b_0^+ + b_0^-) \nonumber \\
%&& - \frac{2\pi T_{D2}}{T_{F1}} \sum_{k_\sigma > 1} \left[\frac{(B^2-R^2)k_\sigma}{B(R^2+B^2)}-\frac{2B}{R^2+B^2}-\frac{1}{2B(k_\sigma-1)}\right] a^+_{k_\sigma}a^-_{-k_\sigma} \nonumber \\
%&&+ \frac{2\pi T_{D2}}{T_{F1}} \sum_{k_\sigma < 1}\left[\frac{(B^2-R^2)k_\sigma}{B(R^2+B^2)}-\frac{2B}{R^2+B^2}-\frac{1}{2B(k_\sigma-1)}\right]  a^-_{-k_\sigma}a^+_{k_\sigma} \nonumber\\
 %&& + \frac{2\pi T_{D2}}{T_{F1}} \sum_{k_\sigma > 1}\left[\frac{k_\sigma}{B}+\frac{1}{2B(k_\sigma-1)}\right] b^-_{-k_\sigma}b^+_{k_\sigma} - \frac{2\pi T_{D2}}{T_{F1}}\sum_{k_\sigma < 1}\left[\frac{k_\sigma}{B}+\frac{1}{2B(k_\sigma-1)}\right] b^+_{k_\sigma}b^-_{-k_\sigma}  \nonumber \\
%&& + \frac{2\pi T_{D2}}{T_{F1}}  \frac{1}{4B} \sum_{k_\sigma \neq \pm 1} \frac{b^+_{k_\sigma}b^+_{-k_\sigma}}{\sqrt{ |1-k^2_\sigma|}}+  \frac{2\pi T_{D2}}{T_{F1}} \frac{1}{4B} \sum_{k_\sigma \neq \pm 1} \frac{b^-_{k_\sigma}b^-_{-k_\sigma}}{\sqrt{ |1-k^2_\sigma|}},\\
\label{angularmnt}
J &=& J^{round}  +  \sqrt{2\pi L_z T_{D2}}  \frac{R}{B} (b_0^+ + b_0^-) \nonumber \\
&& +  \sum_{k_\sigma > 1} \left[\frac{2R^2 k_\sigma}{R^2+B^2}+\frac{2B}{R^2+B^2}+\frac{1}{2(k_\sigma-1)}\right] a^+_{k_\sigma}a^-_{-k_\sigma} \nonumber \\&&- \sum_{k_\sigma < 1}\left[\frac{2R^2 k_\sigma}{R^2+B^2}+\frac{2B}{R^2+B^2}+\frac{1}{2(k_\sigma-1)}\right]  a^-_{-k_\sigma}a^+_{k_\sigma} \nonumber\\
 && +  \sum_{k_\sigma > 1}\frac{ b^-_{-k_\sigma}b^+_{k_\sigma}}{2(k_\sigma -1)} -  \sum_{k_\sigma < 1} \frac{ b^+_{k_\sigma}b^-_{-k_\sigma}}{2(k_\sigma -1)}  \nonumber \\
&& +   \sum_{k_\sigma \neq \pm 1} \frac{b^+_{k_\sigma}b^+_{-k_\sigma}}{4\sqrt{ |1-k^2_\sigma|}}+  \sum_{k_\sigma \neq \pm 1} \frac{b^-_{k_\sigma}b^-_{-k_\sigma}}{4\sqrt{ |1-k^2_\sigma|}} +\sum_{k_\sigma > 0} \frac{2R^2 k_\sigma}{R^2+B^2} a^i_{k_\sigma}a^i_{-k_\sigma} , \\
Q_{D0} &=& Q_{D0}^{round} =  \frac{2\pi L_z T_{D2}}{T_{D0}} B, \cr
\label{chargesdifference} \Delta &:=& Q_{F1} Q_{D0} - J \cr &=&
\sum_{k_\sigma >1} k_\sigma
(b^-_{-k_\sigma}b^+_{k_\sigma}-a^+_{k_\sigma}a^-_{-k_\sigma})-
\sum_{k_\sigma <1} k_\sigma
(b^+_{k_\sigma}b^-_{-k_\sigma}-a^-_{-k_\sigma}a^+_{k_\sigma}) \\ &+&
\sum_{k_\sigma >0} k_\sigma
(b^i_{-k_\sigma}b^i_{k_\sigma}-a^i_{k_\sigma}a^i_{-k_\sigma}) \\
&=& P^{can}_\sigma.
\end{eqnarray}
Here we have chosen to emphasize the Hamiltonian $H$ instead of
the energy $P^0$, though the latter is easily recovered through
the relation (\ref{HBPS}).  Since we have not explicitly included
Fermions, normal ordering has been used to obtain a finite result
for (\ref{opcharges}).   We have also chosen to express the charge
$Q_{F1}$ in terms of $Q_{D0}$ and the angular momentum, as one
sees that the combination $\Delta = Q_{F1} Q_{D0} - J$ defined
above takes a fairly simple form; it is just $P^{can}_\sigma$, the
canonical generator of $\sigma$-translations in our gauge-fixed
theory. A general argument for this form is given in Appendix A,
but we have also verified the result explicitly. As forewarned,
one may further check that $Q_{F1}$ is not the integral of the
electric field momentum $\pi_z$, even at the linear level.

\subsection{Counting States}
\label{entropy}

Let us now fix $H=0$, $Q_{D0}$, and the quantity $\Delta := Q_{F1}
Q_{D0} -J$ (but not $Q_{F1}$ or $J$ individually).  We see from
(\ref{chargesdifference}) that when restricted to BPS states
(those with $\omega =0$), the operator $Q_{F1} Q_{D0} - J$ takes
the form of the energy of a system of 8 right-moving 1+1 massless
scalars. Furthermore, the argument in Appendix A shows that this
follows from general considerations, and thus that the Fermionic
contributions suppressed here must take the corresponding form.
Thus, the entire system is a 1+1 right-moving CFT with central
charge $c=12$. Note that fixing $Q_{D0}$ places no restrictions on
such effective right-moving fields, as $Q_{D0}$ is given by the
magnetic flux, a topological invariant. Thus, the Cardy formula
\cite{Cardy} tells us that the entropy of our system is $S = 2 \pi
\sqrt{2 (Q_{D0} Q_{F1} - J)}$.

What remains is to argue that the entropy depends on the charges
{\it only} through the combination $Q_{D0} Q_{F1} - J$, and to tie
up a loose end having to do with the quantization of charge and
angular momentum.  The latter issue arises from a careful
inspection of (\ref{angularmnt}), which shows that $J$ (and thus
$Q_{F1})$ has a linear term which necessarily leads to a
continuous spectrum. That the spectrum of $Q_{F1}$ is continuous
is an artifact of our not yet imposing that the gauge group is
compact\footnote{The angular momentum $J$ also has continuous spectrum, but it is a familiar result that quantization of $J$ imposes the Dirac quantization condition on the product of electric and magnetic charge.  Note that a proper description of magnetic charge again requires compactification of the gauge group.}.
To do so, we must quotient the configuration space of the
connection by an appropriate translation.  It turns out to be
convenient to deal with both issues simultaneously.

To do so, let us recall that the above quotient compactify the configuration space of the zero mode
$(a_z)_{k=0,k_z=0} = \frac{T_{F1}}{2\pi R L_zT_{D0}} \int dz d
\sigma \ a_z$, where we have chosen the normalization to be such
that $(a_z)_{k=0,k_z=0}$ is compactified
 with period $2\pi$.  Thus, while $(a_z)_{k=0,k_z=0}$ will no longer be a well-defined operator, the exponentiated operator
$e^{in(a_z)_{k=0,k_z=0}}$ will be well-defined for any integer
$n$.

It is useful to consider only the time independent part of this zero mode:
\begin{equation}
(a_z)_{k=0,k_z=0, \omega =0} := \lim_{ T \rightarrow \infty}
\frac{1}{2T}\int_{-T}^T  dt (a_z)_{k=0,k_z=0},
\end{equation}
which depends only on the time independent (and BPS) b-modes. Note
that the exponential $e^{in(a_z)_{k=0,k_z=0, \omega=0}}$ is again
well-defined\footnote{It is also gauge invariant. Invariance under
small diffeomorphisms is manifest from the integration over the
world-volume. Invariance under large diffeomorphisms may be
checked, but in the end is essentially equivalent to the fact
(\ref{transQ}) that the operator translates $Q_{F1}$ by an
integer.  We thank David Gross for raising this issue.} for any
integer $n$.

Now, since $\Pi_z$ is the canonical conjugate to $a_z$ defined by
the action (\ref{D2}), conjugation of $Q_{F1}$ by $e^{in
(a_z)_{k=0,k_z=0, \omega=0}}$ will simply add $n$ units of charge:
\begin{equation}
\label{transQ} e^{-i n (a_z)_{k=0,k_z=0, \omega=0}} Q_{F1}e^{i n
(a_z)_{k=0,k_z=0, \omega=0}} = Q_{F1} + n.
\end{equation}

But we see explicitly that $e^{i n (a_z)_{k=0,k_z=0, \omega=0} }$
commutes with the expression (\ref{chargesdifference}) for
$Q_{D0}Q_{F1}-J$.  Furthermore, since $e^{i n (a_z)_{k=0,k_z=0,
\omega=0} }$ is time-independent, it must commute with $H$ and so
maps BPS states to BPS states. There is thus a unitary (and, in particular,
bijective) map acting within the class of BPS states that changes
$Q_{F1}$, but leaves $Q_{D0}Q_{F1}-J$ invariant.  It follows that
the desired entropy can depend on the charges only through the
combination $Q_{D0}Q_{F1}-J$ and thus that, when all charges are
fixed, the entropy is indeed $2 \pi \sqrt{2 (Q_{D0} Q_{F1} - J)}$ to leading order in the charges.

\subsection{Limits of Validity}
\label{valid}

We have now attained our main goal and verified the conjectured
form of the entropy within the domain of our linearized treatment.
It is important, however, to characterize the size of this domain.
After all, our use of Cardy's formula required $\Delta \gg 1$, and
one might worry that this constraint might be in conflict with our
linear treatment.

We need to estimate the size of some higher order correction to
our calculations.  However, since supertubes are exact solutions
to the Born-Infeld action \cite{MateosTownsend,MateosNgTownsend},
there are no corrections to the solutions at this level.
Furthermore, it has been argued \cite{KMPW} that such supertube
solutions receive no corrections from higher derivative terms in
the D2 effective action\footnote{One may note that T-dualizing
the $O(F^4)$ higher derivative terms obtained in \cite{arkady} would appear
to lead to such corrections. However, since $E=1$ for the supertube one cannot expect the correct behavior to be obtained by considering corrections at any finite order in $F$.  Thus \cite{KMPW} and
\cite{arkady} are not in conflict.  We thank Iosef Bena for this observation.}.  Furthermore, the action vanishes
when evaluated on supertube configurations. Thus, we will not
obtain useful error estimates from the action or equations of
motion.

On the other hand, our charges {\it do} receive corrections from
the higher order terms:  even for supertubes, the expression
(\ref{charges1}) is not quadratic.  Thus we may estimate our
errors by comparing contributions to $Q_{F1}$ from different
orders.  Rather than calculate the third order term, we will
simply compare the second-order contribution with the zero-order
term.  (Note that the linear term gives only a rather trivial
shift of the background and, in particular, is independent of
$\Delta$.)

There are in fact two types of quadratic contributions to
$Q_{F1}$: those appear in $\Delta= Q_{F1}Q_{D0}-J$ and those that
appear in $J$.  Restricting $\Delta$ to be small requires merely
$\Delta \ll Q_{F1}Q_{D0}$.

Let us now consider the quadratic terms in $J$.  We are interested
only in the BPS modes, so we need only include those terms built
from $b^\pm_{k_\sigma}$.  Examination of (\ref{opcharges}) shows
that typical matrix elements of such terms are of rough size
$\sum_{k_\sigma \ge 1} N_{k_\sigma}/k_\sigma,$ where
$N_{k_\sigma}$ is the number operator associated with each mode.
Since $k_\sigma$ takes values in the positive integers, such terms
are always smaller than $\Delta$ and impose no further
restriction.

\section{Discussion}
\label{disc}

We have seen above  that supertubes represent marginal bound
states and that for $Q_{D0}Q_{F1} \gg Q_{D0}Q_{F1} - J \gg 1$ the
entropy of supertube states is given to leading order by $S = 2\pi
\sqrt{2 (Q_{D0}Q_{F1}-J)}$.  This is identical to the
leading-order entropy of all such D0-F1 bound states.  In
particular, once the center-of-mass momenta are fixed we obtain a
discrete set of supertube states despite the presence of an
infinite number of zero-frequency modes. This result is perhaps
most easily explained by noting \cite{KMPW} that $\sigma$ is a
null direction with respect to the (inverse) open-string metric
(defined in \cite{SW}) on the supertube. Thus, our shape degrees
of freedom are more similar to excitations of a 1+1 massless field
than to those of the more familiar sort of zero mode. Note that it
is in fact easier to count states in which all three of $Q_{D0}$,
$Q_{F1}$, and $J$ are fixed than when only $Q_{F1},Q_{D0}$ are
fixed, since restricting $Q_{D0}Q_{F1}-J \ll Q_{F1}Q_{D0}$ allows
us to treat the system perturbatively.

Our results support the conjecture that supertubes provide an
effective description of generic D0-F1 bound states.  It would be
interesting to extend this analysis by applying similar techniques
to the related supergravity solutions of
\cite{EmparanMateosTownsend} or \cite{LuninMaldacenaMaoz}, or
perhaps by studying the linearization around other (less
symmetric) Born-Infeld supertube configurations.  In addition, it
would be of interest to relate our entropy calculations to the
entropy of the two-charge black rings of Emparan and Elvang
\cite{ElvangEmparan}.

We note that results for the multiply wound case where $\tan(X^1/X^2) = \sigma/n$ may also be of interest.  Such results are easily obtained from those above by applying the methods of appendix A and noting that the only change is the replacement $J \rightarrow nJ$ as the tube now rotates $n$
times in the $X^1X^2$ plane under $\sigma \rightarrow \sigma + 2 \pi$. Thus, the entropy of small
fluctuations about the round tube with $n$ wrappings is given by $S=2\pi \sqrt{2(Q_{D0}Q_{F1} -nJ)}$.
For fixed $Q_{D0},Q_{F1},J$ we see that the entropy is greatest for the case $n=1$.

The results above are of use for understanding the two-charge
system, but similar studies for the related three-charge systems
could have implications for black holes and thus be of much
greater interest.  In particular, Mathur et al
\cite{LuninMathur1,LuninMathur2,MathurEssay,Mathur3Q} have
conjectured that similar results hold for such three-charge
systems:  that almost all such bound states can be described in
terms of extended horizon-free configurations in which the entropy
is readily apparent, for example with the distinct states being
labelled by the shape of the object and the values of associated
worldvolume fields.  If this were so, it would leave no room for
black holes as a distinct class of states.  Thus, Mathur et al
wish to conjecture that black holes represent only an effective
statistical average over collections of more fundamental states;
see \cite{LuninMathur1,LuninMathur2,MathurEssay} for details.

Such conjectures cannot be studied using our abelian D2 effective
action, as the third charge in this context corresponds to adding
D4-branes orthogonal to the tube directions.  However, one may
imagine studying linearizations of the three-charge
Dirac-Born-Infeld system of \cite{BK}.  Such calculations are
currently underway in joint work with the authors of \cite{BK}.

Perhaps even more interesting would be to study in detail
linearized fluctuations of the known smooth
\cite{LuninMaldacenaMaoz} D1-D5 supergravity
solutions\footnote{Related studies of metrics with conical
deficits were begun in \cite{Mathur3Q} and are being continued by
those authors.}. However, merely adding momentum to such solutions
seems unlikely to yield enough states to account for the full
entropy of the three-charge system.  In particular, in order to
find enough entropy it is not sufficient simply to find BPS modes
with arbitrarily $k_z$. Instead, recall that the 3-charge entropy
involves the product $Q_{D1}Q_{D5}P$, whereas simply adding a
$1+1$ field theory in the $z-t$ plane would at most add a function
of $P$ to the 2-charge entropy.  Thus, it would appear that one
would need to find BPS modes with $k_\sigma$ and $k_z$ both
arbitrary. In other words, the tube must not only support
travelling waves, but must support independent travelling waves at
each value of $\sigma$. Such a result cannot be obtained from any
non-degenerate quadratic 2+1 dispersion relation, though we cannot
immediately rule out the possibility that it might arise due to
complicated linear couplings between the various degrees of
freedom.   We understand that related issues are currently being
explored by Mathur and collaborators (in an extension of
\cite{Mathur3Q}). Working from a different perspective, we also
hope to report related further results (with our collaborators) in
the near future.  Finally, it remains possible that other
yet-unsuspected generalizations of known solutions will lead to
enough states to account for the full entropy of the three-charge
system.

\medskip

{\bf Acknowledgments:} The authors would like to thank Henriette
Elvang, Steve Giddings, Gary Horowitz, Samir Mathur, and Joe
Polchinski for thought-provoking conversations.  We particularly
thank David Mateos and Iosef Bena for several useful discussions.
This work was supported in part by NSF grants PHY00-98747,
PHY99-07949, and PHY03-42749, by funds from Syracuse University,
and by funds from the University of California. BCP was supported
in part by a Syracuse University Graduate Fellowship. In addition,
BCP thanks UCSB for its hospitality during the latter part of this
work.

\appendix

\section{Gauge Fixing and Charges}

In this appendix we show how the important relations (\ref{HBPS}) and (\ref{chargesdifference}) follow directly from general considerations of symmetries and our gauge fixing scheme.  As a result, they represent a useful check on our calculations.

It will be helpful to distinguish here between the full Dirac-Born-Infeld  Lagrangian of (\ref{D2}), which we
denote $L$, and the quadratic gauge fixed Lagrangian ($L^{(2)}_{gf}$) explicitly displayed in (\ref{action}).
We remind the reader that $L^{(2)}_{gf}$ is obtained from $L$ in two stages, first gauge fixing $L$
to form $L_{gf}$, and then taking the quadratic term which yields $L^{(2)}_{gf}$.  In particular,
note that passing to $L^{(2)}_{gf}$ discards the constant term corresponding to evaluating $L$ on our background, as this term  is of order zero in our perturbations.

In fact, we argue in somewhat more generality below.  Let us consider the Lagrangian $\tilde L_{gf}$ which differs from $L_{gf}$ only by subtracting the background value, while retaining all of the higher terms:
\begin{equation}
\label{tildeL}
\tilde L_{gf} := L_{gf} - L|_{Background} = L^{(2)}_{gf} + \  {\rm higher \ order \  terms}.
\end{equation}
We begin by noting that invariance under $t$ and $\sigma$
reparametrizations implies two important identities for $L$, which
we may call the Hamiltonian and momentum constraints:

\begin{eqnarray}
\sum_\mu \frac{\partial L}{\partial (\partial_t X^\mu)} \partial_t
X^\mu + \sum_i \frac{\partial L}{\partial (\partial_t A_i)}
 F_{ti} &=& L \label{tid} \\
 \sum_\mu \frac{\partial
L}{\partial (\partial_t X^\mu)} \partial_\sigma X^\mu + \sum_i
\frac{\partial L}{\partial (\partial_t A_i)} F_{\sigma i} &=&  0 .
\label{sid}
\end{eqnarray}

We now use the first of these results to identify $H$ in terms of
$P^0, Q_{F1}$ and $Q_{D0}$. The Hamiltonian $H$ is by definition

\begin{eqnarray}
H &=& \int dz d\sigma \left( \frac{\partial \tilde L_{gf}}{\partial (\partial_t \eta^\mu)} \partial_t \eta^\mu + \sum_i \frac{\partial \tilde L_{gf}}{\partial (\partial_t a_i)} \partial_t a_i - \tilde L_{gf} \right) \cr
&=& \int dz d\sigma \left( \sum_{\mu} \frac{\partial L_{gf}}{\partial (\partial_t X^\mu)} \partial_t X^\mu + \sum_i \frac{\partial L_{gf}}{\partial (\partial_t A_i)} \partial_t A_i - {\rm sgn}(E) \Pi_z  - L_{gf}  + L|_{\rm Background}  \right) \nonumber ,
\end{eqnarray}
where we have used (\ref{tildeL}) and the fact that the only
time-dependent background field not completely fixed by the gauge
condition is $A_z$, whose time derivative is $E=\pm1$ and whose
conjugate momentum defined from (\ref{D2}) is $\Pi_z$. Recall that
canonical transformations do not affect the Hamiltonian, so that
we may consistently ignore the extra integration by parts
mentioned below (\ref{action}) which would replace $\Pi_z$ by
$\pi_z$ and perform a compensating change in $L_{gf}$.

Now, $L_{gf}$ is obtained from $L$ by imposing the requirements
$X^0=t$, $X^3=z$, $X^1=R(t,z,\sigma) \cos{\sigma}$,
$X^2=R(t,z,\sigma) \sin{\sigma}$ and $A_0=0$.  We denote this
process by $|_{gf}$, e.g. $L_{gf} =  L |_{gf}$.  Expressing $H$ in
terms of $L$, we find
\begin{eqnarray}
H &=& \int dz d\sigma \left( \sum_{\mu }  \frac{\partial L}{\partial (\partial_t X^\mu)} \partial_t X^\mu
-  \frac{\partial L}{\partial (\partial_t X^0)}
+ \sum_i \frac{\partial L}{\partial (\partial_t A_i)} \partial_t A_i - {\rm sgn}(E) \Pi_z  - L + L|_{\rm Background}  \right)|_{gf}, \cr
&=&
- \int dz d\sigma \left(
\frac{\partial L}{\partial (\partial_t X^0)}
+  {\rm sgn}(E) \Pi_z  -  L|_{\rm Background}  \right)|_{gf},
\end{eqnarray}
where in the last line we have used the Hamiltonian constraint
(\ref{tid}).

Finally, the general form of the Dirac-Born-Infeld action implies the relation
\begin{equation}
\frac{\partial L}{\partial G_{00}}|_{G=\eta} = - \frac{1}{2}
\sum_a \frac{\partial L}{\partial (\partial_a X^0)} \partial_a
X^0,
\end{equation}
where $|_{G=\eta}$ denotes that we evaluate the expression (after taking the derivative) for the special case where $G_{ab}$ is the Minkowski metric.  After gauge fixing this becomes
\begin{equation}
T_{00}|_{gf,G=\eta} =  2 \frac{\partial L}{\partial
G_{00}}\bigg|_{gf,G=\eta} =  - \frac{\partial L}{\partial
(\partial_t X^0)}  |_{gf,G=\eta}.
\end{equation}
Using this together with $L_{Background}  = - B{\rm sgn}(B)  $
and the definition of the charges (\ref{charges0}), (\ref{charges1})
we find
\begin{eqnarray}
H &=&  \int dz d\sigma \left( T_{00} - {\rm sgn}(E) \Pi_z - B {\rm sgn}(B)
\right)|_{gf}\cr
&=& P^0 - |Q_{F1}| L_z T_{F1}  - |Q_{D0}| T_{D0},
\label{Hresult}
\end{eqnarray}
where in the final step we have used the fact that the integrated magnetic flux is a
topological invariant and so is always given by
its value in the round tube background. Again we emphasize that the validity of (\ref{Hresult}) is in
no way restricted to the linear approximation.
Note that the main text
primarily studies the case ${\rm sgn}(E)={\rm sgn}(B)=1$ for which
$Q_{F1},Q_{D0} >0$.

Now, $H$ is the generator of time translations in the quadratic
gauge-fixed theory.  We may of course also consider
$P^{can}_\sigma$, the generator of $\sigma$-translations in the
quadratic gauge-fixed theory. We apply the analogous reasoning to
$P^{can}_\sigma$, which by definition takes the form
\begin{eqnarray}  P_{\sigma} &=& \int dz d \sigma
\frac{
\partial \tilde{L}_{gf}}{\partial (\partial_t \eta^{\mu})}
\partial_{\sigma} \eta^{\mu} + \frac{ \partial
\tilde{L}_{gf}}{\partial (\partial_t a_i)} (\partial_{\sigma} a_i -
\partial_i a_{\sigma}) \nonumber \\ \label{momenta}
 &=&\int dz d \sigma  \sum_{\i=3 }^{8}  \frac{
\partial L_{gf}}{\partial (\partial_t X^{i})} \partial_{\sigma} X^{i}
+\frac{
\partial L_{gf}}{\partial (\partial_t R) } \partial_{\sigma}
R + \frac{\partial L_{gf}}{\partial (\partial_t A_z)}
(\partial_{\sigma} A_z - \partial_z A_{\sigma} + B).
\end{eqnarray}
Let us now compute,
\begin{eqnarray} & & \frac {\partial
L}{\partial
(\partial_t X^1)} \partial_{\sigma} X^1|_{gf} +\frac {\partial
L}{\partial (\partial_t X^2)} \partial_{\sigma} X^2
|_{gf} \nonumber \\
&=& \partial_{\sigma} R (\cos{ \sigma} \frac{\partial L}{\partial
(\partial_t X^1)}|_{gf}  + \sin{ \sigma} \frac{\partial L}{\partial
(\partial_t X^2)}|_{gf})  +
(R \cos{\sigma} \frac{\partial L}{\partial (\partial_t X^2)}|_{gf} -
 R \sin{\sigma} \frac{\partial L}{\partial (\partial_t X^1)}|_{gf}) \nonumber \\
&=& \partial_{\sigma} R \frac{\partial L_{gf}}{\partial (\partial_t
R)} + \mathcal{L}_{12} , \end{eqnarray} where $\mathcal{L}_{12}$ is
the density along the $S^1$ of the component of angular momentum (which we have called $J$) associated with the $X^1X^2$ plane. Substituting
the above expression in (\ref{momenta}),
\begin{eqnarray} P_{\sigma} & =&  \int dz d \sigma  \left( \frac{
\partial L}{\partial (\partial_t X^{\mu})} \partial_{\sigma}
X^{\mu} +  \frac{
\partial L}{\partial (\partial_t A_z)} F_{\sigma z} \right) |_{gf} + \Pi_z B
-\mathcal{L}_{12},
 \end{eqnarray}
and using the identity (\ref{sid}), one arrives at the relation
\begin{equation}
P^{can}_\sigma = Q_{F1} Q_{D0} - J.
\end{equation}
We note that the form of (\ref{chargesdifference}) then follows
immediately from our identification of the creation and
annihilation operators.

\section{Quadratic Expansions}

This appendix simply lists the formulae, suppressed in the main
text, which describe 1) the quadratic expansions of the action for
the $z$-independent fields and 2) the charges in terms of the
perturbations $\eta^\mu, a_i$.

The action for the $z$-independent modes is
\begin{eqnarray}
S &=& S_{round} + S^{(2)} + {\rm \ higher \ order \ terms} \\
 S^{(2)} &=&  -  {\rm sgn}(B) \frac{L_zT_{D2}}{2} \int dt d\sigma \Bigl[ -\frac{R^2+B^2}{B}((\partial_t r)^2 + |\partial_t \eta|^2) \nonumber  \\
&& - 2{\rm sgn}(E)(\partial_t r \partial_{\sigma} r + \partial_t \eta^i \partial_{\sigma} \eta^i)
-{\rm sgn}(E)\frac{2R}{B}(\partial_t a_z r- \partial_t r a_z) \nonumber \\
 &-&\frac{R^2(R^2+B^2_0)}{B^3} ( \partial_t a_z)^2   - \frac{1}{B} (\partial_t a_\sigma)^2 -  {\rm sgn}(E)\frac{2R^2}{B^2} \partial_t a_z \partial_\sigma a_z   \label{action}\Bigr] .
\end{eqnarray}
In computing (\ref{action}) from (\ref{D2}) we have performed an integration by parts, which induces
a canonical transformation designed to make the momenta (\ref{momenta}) take a more symmetric form.   As a result,
the canonical momentum $\pi_z$, defined by the action (\ref{action}) conjugate to the connection to differ by linear terms from the $\Pi_z$ (\ref{charges1}), defined by (\ref{D2}).  Thus, while the electric charge $Q_{F1}$ remains the integral of $\Pi_z$, it is not the integral of $\pi_z$.

The charges and Hamiltonian take the form
\begin{eqnarray}
\label{B2}
Q_{D0} &=& Q_{D0}^{round},\\
Q_{F1} &=& \frac{1}{T_{F1}}   \int d\sigma \Pi_{z}  \nonumber \\
&=& Q_{F1}^{round}+  {\rm sgn}(EB) \frac{T_{D2}}{2T_{F1}}   \int d\sigma \ \Bigl[ \frac{4R}{B} r + 2 {\rm sgn}(E) \frac{R^2(R^2+B^2)}{B^3}  e_z  \nonumber \\
&&  +  \frac{R^2(R^2+B^2)}{B^3}((\partial_t r)^2 + |\partial_t \eta|^2) \nonumber  \\
&& - \frac{R^4}{B^3}((\partial_z r)^2 + |\partial_z \eta|^2)  \nonumber \\
&&+ \frac{2}{B}(r^2 +(\partial_{\sigma} r)^2 +|\partial_{\sigma} \eta|^2) + {\rm sgn}(E) \frac{2(R^2+B^2)}{B^2}(\partial_t r \partial_{\sigma}r + \partial_t \eta^i \partial_{\sigma} \eta^i) \nonumber \\
 &&  +\frac{3R^4(R^2+B^2)}{B^5} e^2_z +\frac{2 R^2}{B^3} b^2 + \frac{R^2}{B^3} e^2_{\sigma} \nonumber \\
&& - {\rm sgn}(E)\frac{2R^2(3R^2+B^2)}{B^4}  e_z b -\frac{4R}{B^2} rb  +{\rm sgn}(E) \frac{4R(2R^2+B^2)}{B^3}r e_z
\Bigr] ,\\
J &=& J^{round} +{\rm sgn}(B)\frac{T_{D2}}{2} \int dz  d\sigma \ \Bigl[  {\rm sgn}(E)4Rr+2 \frac{R^2(R^2+B^2)}{B^2}e_z  \nonumber  \\
&& + {\rm sgn}(E)2 r^2 + {\rm sgn}(E) \frac{R^2(R^2+B^2)}{B^2} ((\partial_t r)^2 + |\partial_t \eta|^2) \nonumber \\
&& - \frac{ R^4}{B^2}((\partial_z r)^2 + |\partial_z \eta|^2) + {\rm sgn}(E) \frac{3R^4(R^2+B^2)}{B^4} e_z^2 \nonumber \\
&& +\frac{R^4}{B^2} e_{\sigma}^2-\frac{4R^4}{B^3} e_z b +\frac{4R(2R^2+B^2)}{B^2}re_z
\Bigr] ,\\
P^0 &=& P^0_{round} +\frac{{\rm sgn}(B)} {2} T_{D2} \int dz d\sigma \Bigl[\frac{4R}{B} r + 2 {\rm sgn}(E)\frac{R^2(R^2+B^2)}{B^3} e_z \nonumber \\
&& + \frac{(R^2+B^2)^2}{B^3}((\partial_t r)^2 + |\partial_t \eta|^2)  - \frac{R^2(R^2-B^2)}{B^3}((\partial_z r)^2 + |\partial_z \eta|^2)  \nonumber \\
&& + \frac{2}{B}(r^2 +(\partial_{\sigma} r)^2 +|\partial_{\sigma} \eta|^2) +{\rm sgn}(E)\frac{2(R^2+B^2)}{B^2}(\partial_t r \partial_{\sigma}r + \partial_t \eta^i \partial_{\sigma} \eta^i) \nonumber \\
&& +\frac{R^2(3R^2+B^2)(R^2+B^2)}{B^5} e^2_z +\frac{2 R^2}{B^3} b^2 + \frac{R^2+B^2}{B^3} e^2_{\sigma} \nonumber \\
&& -{\rm sgn}(E)\frac{2R^2(3R^2+B^2)}{B^4}  e_z b -\frac{4R}{B^2} rb  +{\rm sgn}(E)\frac{4R(2R^2+B^2)}{B^3}r e_z \Bigr] , \\
H &=& P^0 - |Q_{D0}| T_{D0}  - |Q_{F1}| T_{F1}L_z , \nonumber \\
&=&{\rm sgn}(B) \frac{T_{D2}}{2}  \int dz d \sigma \ \Bigl[\frac{R^2+B^2}{B}((\partial_t r)^2+|\partial_t \eta|^2)+\frac{R^2}{B}((\partial_z r)^2+|\partial_z \eta|^2)  \nonumber \\
&&  \ \ \ \ \ \ \ \ \ \ \ \ \ \ \ \ \ \ + \frac{R^2(R^2+B^2)}{B^3}(\partial_t a_z)^2 + \frac{1}{B}(\partial_t a_\sigma)^2 \Bigr].
\label{B6}
\end{eqnarray}
Note in particular that $H$ is not the energy $P^0$ that couples
to the gravitational field.  Instead, $H$ measures the extent to which a state is excited above
the BPS bound.  Note also that expressions (\ref{B2}-\ref{B6}) are
valid even when the fields depend on $z$.

\end{document}